\documentclass[sn-nature]{sn-jnl}% Basic Springer Nature Reference Style/Chemistry Reference Style

\usepackage{graphicx}%
\usepackage{multirow}%
\usepackage{amsmath,amssymb,amsfonts}%
\usepackage{amsthm}%
\usepackage{mathrsfs}%
\usepackage[title]{appendix}%
\usepackage{xcolor}%
\usepackage{textcomp}%
\usepackage{manyfoot}%
\usepackage{booktabs}%
\usepackage{algorithm}%
\usepackage{algorithmicx}%
\usepackage{algpseudocode}%
\usepackage{listings}%
\usepackage{setspace}
\usepackage{xr}
\usepackage{hyperref} % For clickable links
\usepackage{xr-hyper}
\usepackage{siunitx}

\theoremstyle{thmstyleone}%
\theoremstyle{thmstyletwo}%

\theoremstyle{thmstylethree}%
\begin{document}

\title[Article Title]{A Topological Superconductor Tuned by Electronic Correlations}

%Tuning Topological Superconductivity via Electronic Correlation
%Interplay of Strong Correlation and Topology in $\text{FeTe}_{\text{1-x}}\text{Se}_{\text{x}}$ Thin Film}

%%\equalcont{These authors contributed equally to this work.}

% \author{\fnm{UChicago-UIUC-WVU} \sur{Collaboration}}

\author[1]{\fnm{Haoran} \sur{Lin}}

\author[2]{\fnm{Christopher L.} \sur{Jacobs}}

\author[1]{\fnm{Chenhui} \sur{Yan}}

\author[3]{\fnm{Gillian M.} \sur{Nolan}}

\author[1]{\fnm{Gabriele} \sur{Berruto}}

\author[1]{\fnm{Patrick} \sur{Singleton}}

\author[1]{\fnm{Khanh Duy} \sur{Nguyen}}

\author[1]{\fnm{Yunhe} \sur{Bai}}

\author[1]{\fnm{Qiang} \sur{Gao}}

\author[4]{\fnm{Xianxin} \sur{Wu}}

% \author[5]{\fnm{Yan} \sur{Li}}

% \author[5]{\fnm{Hao} \sur{Zheng}}

% \author[5]{\fnm{Hua} \sur{Zhou}}

% \author[5]{\fnm{Zhan} \sur{Zhang}}

\author[5]{\fnm{Chao-Xing} \sur{Liu}}

\author[1]{\fnm{Gangbin} \sur{Yan}}

\author[1]{\fnm{Suin} \sur{Choi}}

\author[1]{\fnm{Chong} \sur{Liu}}

% \author[5]{\fnm{Yue} \sur{Cao}}

\author[6]{\fnm{Nathan P.} \sur{Guisinger}}

\author[3]{\fnm{Pinshane Y.} \sur{Huang}}

\author[2]{\fnm{Subhasish} \sur{Mandal}}

\author*[1]{\fnm{Shuolong} \sur{Yang}}\email{yangsl@uchicago.edu}

\affil[1]{\orgdiv{Pritzker School of Molecular Engineering}, \orgname{The University of Chicago}, \orgaddress{\city{Chicago}, \postcode{60637}, \state{IL}, \country{USA}}}

\affil[2]{\orgdiv{Department of Physics and Astronomy}, \orgname{West Virginia University}, \orgaddress{\city{Morgantown}, \postcode{26506}, \state{WV}, \country{USA}}}

\affil[3]{\orgdiv{Department of Materials Science and Engineering}, \orgname{University of Illinois Urbana-Champaign}, \orgaddress{\city{Urbana}, \postcode{61801}, \state{IL}, \country{USA}}}

\affil[4]{\orgdiv{Institute of Theoretical Physics}, \orgname{Chinese Academy of Sciences}, \orgaddress{\city{Beijing}, \postcode{100190},  \country{P.R.China}}}

\affil[5]{\orgdiv{Department of Physics}, \orgname{Pennsylvania State University}, \orgaddress{\city{University Park}, \postcode{16802}, \state{PA}, \country{USA}}}

\affil[6]{\orgname{Argonne National Laboratory}, \orgaddress{\city{Lemont}, \postcode{60439}, \state{IL}, \country{USA}}}


\abstract{A topological superconductor, characterized by either a chiral order parameter~\cite{schnyder2008classification, kitaev2009periodic} or a chiral topological surface state in proximity to bulk superconductivity~\cite{fu2008superconducting}, is foundational to topological quantum computing. As in other topological phases of matter~\cite{cao2018correlated,lu2024fractional,cai2023signatures,park2023observation}, electronic correlations can tune topological superconductivity via modifications of the low-energy Fermiology~\cite{qi2010topological}. Such tuning has not been realized so far. Here we uncover a unique topological superconducting phase in competition with electronic correlations in 10-unit-cell thick FeTe$_{x}$Se$_{1-x}$ films grown on SrTiO$_{3}$ substrates. When the Te content $x$ exceeds $0.7$, we observe a rapid increase of the effective mass for the Fe $d_{xy}$ band, with the emergence of a superconducting topological surface state confirmed by high-resolution angle-resolved photoemission spectroscopy; however, near the FeTe limit, the system enters an incoherent regime where the topological surface state becomes unidentifiable and superconductivity is suppressed. Theory suggests that the electron-electron interactions in the odd-parity $xy^-$ band with a strong $d_{xy}$ character lead to an orbital-selective correlated phase. Our work establishes FeTe$_{x}$Se$_{1-x}$ thin films as a unique platform where electronic correlations sensitively modulate topological superconductivity, suggesting opportunities to use tunable electron-electron interactions to engineer new topological phases in a broad class of materials.
}




\maketitle

\section{Introduction}\label{sec1}
% \textcolor{red}{[Haoran: rewrote introduction.]}
A topological superconductor enables non-abelian statistics in topological quantum architectures~\cite{lian_topological_2018}. Initial theoretical predictions of topological superconductors focused on those with a $p_{x} + ip_{y}$ order parameter~\cite{kitaev_unpaired_2001}, yet a formally equivalent scheme is a topological insulator (TI) proximity-coupled with an $s$-wave superconductor~\cite{fu2008superconducting}. FeTe$_{x}$Se$_{1-x}$ (FTS) thus attracts broad interests owing to its intrinsic topological surface state (TSS) proximity-coupled with the material's own bulk superconductivity. This topological superconducting phase was predicted for FTS with Te content $x$ exceeding $0.5$~\cite{wang_topological_2015, wu_topological_2016, xu_topological_2016}, and first demonstrated in bulk FeTe$_{0.55}$Se$_{0.45}$~\cite{zhang_observation_2018}. Meanwhile, strong electronic correlations also emerge when systematically tuning $x$~\cite{huang_correlation-driven_2022, li2024orbital}. It remains experimentally unknown how strong correlations impact topological superconductivity in FTS. Understanding this aspect is particularly crucial for ultrathin films of FTS, which are key to enabling quantum device fabrication. %for its simultaneous display of strong electronic correlations, non-trivial band topology, and superconductivity, making it a promising platform for studying topological superconductivity. In bulk single crystal FeTe$_{0.55}$Se$_{0.45}$, intrinsic topological surface states (TSSs) and surface superconductivity have been demonstrated~\cite{zhang_observation_2018}, but direct evidence of TSS in thin-film FTS has remained elusive. 

\section{Results}\label{sec2}
\begin{figure}[h]%
\centering
\includegraphics[width=1\textwidth]{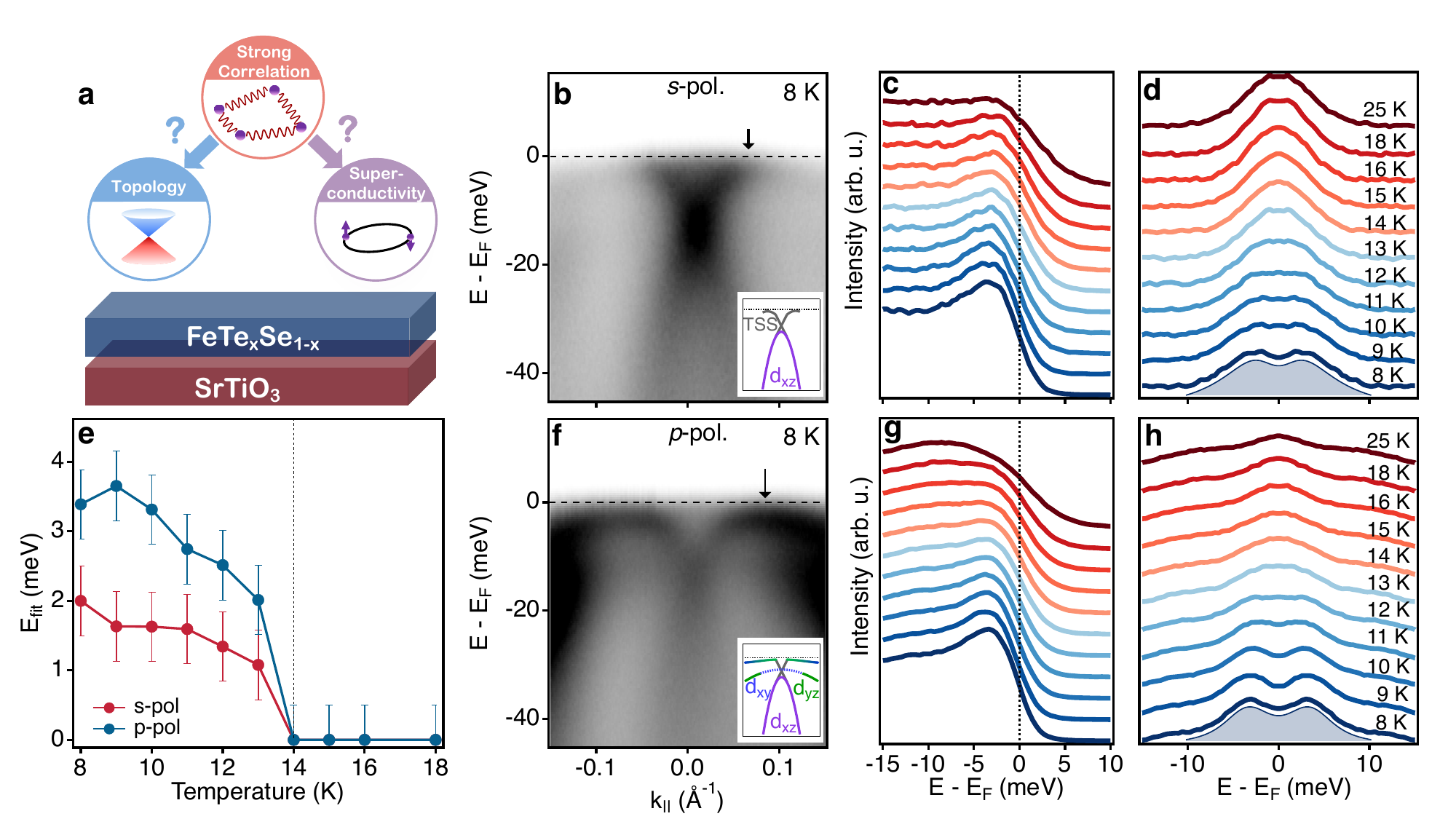}
\caption{\textbf{Superconducting gap measurements on 10 UC FeTe$_{0.93}$Se$_{0.07}$. a,} Schematic illustration of the goal of this study: how strong electronic correlations impact band topology and superconductivity in thin-film FeTe$_x$Se$_{1-x}$. \textbf{b,} ARPES spectrum on 10 UC FeTe$_{0.93}$Se$_{0.07}$ measured at 8 K along $\overline{\Gamma}-\overline{\text{X}}$ using $s$-polarized 6 eV photons. The black arrow indicates the momentum at which energy distribution curves (EDCs) are taken. Inset shows the schematic of the polarization-selected band structure. \textbf{c,} Selected EDCs at various temperatures. \textbf{d,} EDCs symmetrized around the zero energy. \textbf{e,}  Temperature dependence of the extracted superconducting gap $E_{\text{fit}}$ for the two polarization schemes, determined from the symmetrized EDCs in \textbf{d} and \textbf{h}. Above $14$~K, the quality of fits using either a single- or double-Lorentzian model becomes indistinguishable, indicating that the gap has effectively closed (Supplementary Fig.~7). Within this framework, the gap extracted at $14$~K using the established Norman model~\cite{norman_phenomenology_1998} should also be indistinguishable from zero. We thus define our effective error bars based on this criterion. \textbf{f,g,h}: Same as \textbf{b,c,d}, but using $p$-polarized 6 eV photons.}
\label{fig1}
\end{figure}

\begin{figure}[h]%
\centering
\includegraphics[width=1\textwidth]{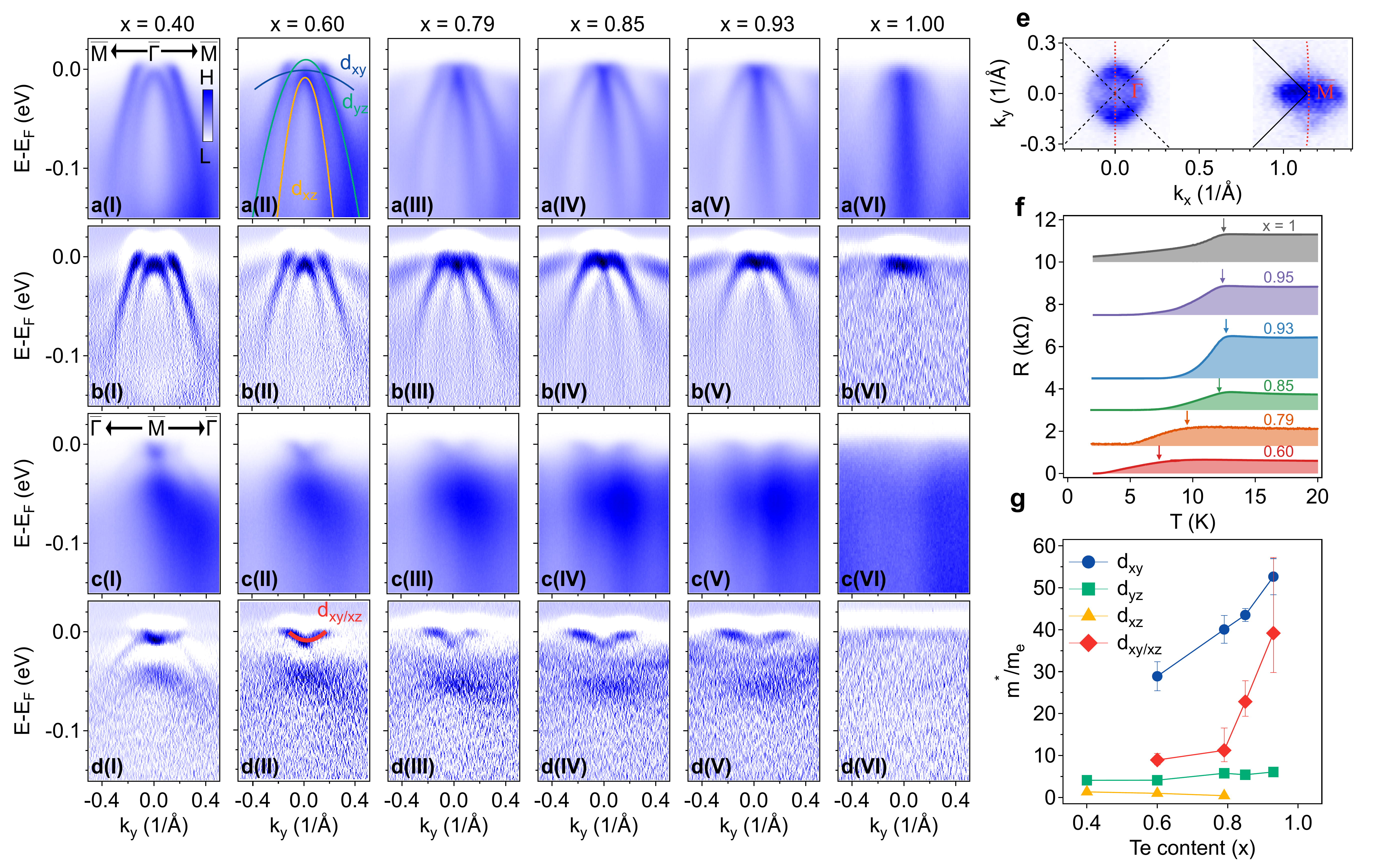}
\caption{\textbf{Experimental spectral functions of 10 UC FeTe$_{x}$Se$_{1-x}$ films measured by ARPES using $21.2$~eV photons. a(I)-a(VI),} Evolution of the ARPES spectra as a function of the Te content ($x$) near the $\overline{\Gamma}$ point. \textbf{b(I)-b(VI),} Corresponding 2D curvature plots of the raw data in \textbf{a}, extracted using the protocol reported by Ref.~\cite{zhang_precise_2011}. \textbf{c(I)-c(VI),} Evolution of the ARPES spectra near the $\overline{\text{M}}$ point. \textbf{d(I)-d(VI),} Corresponding 2D curvature plots of the raw data in \textbf{c}. \textbf{e,} The Fermi surface map of 10 UC $\text{FeTe}_{\text{0.40}}\text{Se}_{\text{0.60}}$. Red dashed lines represent the momentum-space trajectories for the ARPES data shown in \textbf{a} and \textbf{c}. \textbf{f,} Resistance-vs-temperature curves for FTS thin films. The resistance values of each curve have been vertically offset for clarity. \textbf{g,} Extracted effective masses normalized by the free-electron mass for the $d_{xz}$, $d_{yz}$ and $d_{xy}$ bands near $\overline{\Gamma}$, as well as the hybridized $d_{xy/xz}$ band near $\overline{\text{M}}$. Error bars indicate one-standard-deviation ($1\sigma$) uncertainties of the fitting results. All ARPES measurements were done at a sample temperature of $20$~K.}\label{fig2}
\end{figure}

We begin our investigation by showing evidence of topological superconductivity in thin-film FTS: the opening of a superconducting gap on the TSS in a 10-unit-cell thick (10 UC) FeTe$_{0.93}$Se$_{0.07}$ film grown on SrTiO$_{3}$ (STO) (Fig.~\ref{fig1}). This thickness is chosen to minimize strain relaxation from the STO substrate while still preserving a quasi-3D band structure. The Te content $x$ is determined by considering both the $c$-axis lattice constant and the Te:Se flux ratio, cross-checked by inductively coupled plasma mass spectrometry (Supplementary Note 1). Based on matrix element calculations for angle-resolved photoemission spectroscopy (ARPES) (Supplementary Note 2), we use $s$-polarized light to enhance sensitivity to the $d_{xz}$ orbital, and $p$-polarized light for the $d_{yz}$, $d_{xy}$, and $p_{z}$ orbitals. All of these orbitals contribute to the TSS which is probed by both polarizations (Fig.~\ref{fig1}\textbf{b},\textbf{f}). Energy distribution curves (EDCs) taken at the Fermi momentum ($k_{\rm F}$) of the TSS exhibit an enhanced quasiparticle peak when temperature decreases (Fig.~\ref{fig1}\textbf{c}). We symmetrize the EDCs about the Fermi level ($E_{\rm F}$) determined by polycrystalline gold, and observe a gap opening below $\sim 13$~K (Fig.~\ref{fig1}\textbf{d}). The EDC symmetrization is justified by the fact that $13$~K indeed corresponds to the superconducting onset temperature ($T_{\rm c}^{\rm onset}$) observed in electrical transport (Fig.~\ref{fig2}\textbf{f}), and that superconducting gaps are particle-hole symmetric. The band structure revealed by the $p$-polarized light (\textit{the $p$-spectrum}) is more complex (Fig.~\ref{fig1}\textbf{f}). Due to the opening of superconducting gaps, the back-bending sections of the TSS and the $d_{xy}$/$d_{yz}$ hybridized band connect with each other (Supplementary Note 3), resulting in a combined quasi-flat band consistent with previous measurements on bulk FeTe$_{0.55}$Se$_{0.45}$~\cite{jia2024absence}. Here, the spectral gap opens over an extended momentum range beyond $k_{\rm F}$'s of individual bands. We extract EDCs at the momentum corresponding to the smallest gap value (Fig.~\ref{fig1}\textbf{f}). We fit the symmetrized EDCs using the established Norman model~\cite{norman_phenomenology_1998} and extract the superconducting gaps for both light polarizations (Fig.~\ref{fig1}\textbf{e}). Above $\sim 14$~K the experimental data does not allow a robust fitting as the superconducting gap vanishes (Supplementary Fig.~7). Importantly, the gap extracted from the $p$-spectrum is on average $90$\% higher than that from the $s$-spectrum. This is because the bulk $d_{xy}$/$d_{yz}$ band has a strong contribution to the flat-band feature in the $p$-spectrum (Fig.~\ref{fig1}\textbf{f}), whereas the gap extracted from the $s$-spectrum reflects the superconducting order parameter on the TSS. The gap on the TSS is smaller due to proximity coupling with bulk superconductivity. Our results directly support the existence of topological superconductivity and superconducting proximity coupling in multilayer FTS films.

Having verified the superconducting gap on the TSS in thin-film FTS, we now perform a systematic investigation to reveal how varying the Te content ($x$) modifies the electronic structure and influences topological superconductivity. Here in this study, we identify a unique superconducting phase where the electrons near $E_{\rm F}$ are strongly correlated and nearly localized in 10 UC FTS grown on STO. This results from a complex interplay between strong correlations, nontrivial topology, and superconductivity. When $x$ crosses $0.7$, a topological surface state (TSS) emerges and persists up to $x = 0.95$. When $x$ approaches 1, the orbital-selective strong correlations drive the $d_{xy}$ electrons to be localized, leading to an unidentifiable TSS and an incomplete superconducting transition down to $2$~K. Our combined ARPES and DMFT results unveil that the many-body physics of the $d_{xy}$ orbital sensitively controls topological superconductivity through the $xy^-$/$d_{xz}$ band inversion~\cite{kim2024orbital}. These results allow us to elucidate the phase diagram of FTS multilayer films.

%The evolution of the band renormalization in 10UC FTS thin films at $7.5$~K is revealed by ARPES using $21.2$~eV photons. 

%\textcolor{red}{A representative, atomic-resolution STEM image taken from 10UC FeTe$_{0.94}$Se$_{0.06}$ is plotted in Fig.~\ref{fig1}\textbf{c}, demonstrating no in-plane strain relaxation away from the interface. The variation of the in-plane lattice constant is within the statistical uncertainties (Fig~\ref{fig1}\textbf{d}). Moreover, the average chalcogen heights above and below the Fe sheets are $1.83\pm 0.08$ and $1.66\pm 0.05$~\AA, respectively. These values are 4\% higher and 9\% lower than the counterpart of bulk FeTe~\cite{viennois_effect_2010}, and have a substantial impact on the electronic and magnetic properties which we will discuss later.}

% We grow 10 UC FTS films on STO substrates with a systematically increased $x$ using molecular beam epitaxy (MBE). This thickness is chosen to minimize strain relaxation from the STO substrate while still preserving a quasi-3D band structure. The Te content $x$ is determined by considering both the $c$-axis lattice constant and the Te:Se flux ratio (Supplementary Note 1). 

Fig.~\ref{fig2}\textbf{a(I)-a(VI)} displays the electronic structures measured with $21.2$~eV photons at the projected Brillouin zone center ($\overline{\Gamma}$), where three hole-like bands are identified. From the innermost to the outermost, these bands are attributed primarily to the $d_{xz}$, $d_{yz}$, and $d_{xy}$ orbital characters (Fig.~\ref{fig2}\textbf{a(II)}), respectively, based on the light-polarization-dependent matrix elements (Supplementary Note 2, Supplementary Table 1). These assignments are consistent with the previous results on bulk FTS~\cite{yi_observation_2015, liu_experimental_2015, huang_correlation-driven_2022}. Meanwhile, spin-orbit coupling (SOC) leads to hybridization of the Fe $3d$ orbitals (Supplementary Note 5). Here, we follow the convention of referring to the Fe $3d$ bands using their dominant orbital characters~\cite{yi_observation_2015, liu_experimental_2015, huang_correlation-driven_2022}. %\textcolor{red}{Our polarization-dependent 6 eV ARPES data (Supplementary Note 5) further verify these assignments.} 
Local minima of the energy distribution curves (EDCs) and the momentum distribution curves (MDCs) in the 2D curvature plots (Fig.~\ref{fig2}\textbf{b(I)-b(VI)}) are extracted to obtain the band energies at different momenta (Supplementary Fig.~10). The band dispersions are subsequently fitted to parabolic functions to obtain the effective masses, plotted in Fig.~\ref{fig2}\textbf{g}. From x = 0.60 to x = 0.93, the effective mass of the $d_{xy}$ band increases from $29 \pm 3$~$m_{e}$ to $53 \pm 4$~$m_{e}$, where $m_{e}$ is the free electron mass; the counterpart for the $d_{yz}$ band increases from $4.1 \pm 0.4$~$m_{e}$ to $6.1 \pm 0.3$~$m_{e}$. Importantly, the spectral weight of the $d_{xy}$ band in the raw ARPES data is heavily suppressed in FeTe (Fig.~\ref{fig2}\textbf{a(VI)}). The $d_{xz}$ band dispersion is parabolic for $x\leq 0.60$, and becomes quasi-linear near $E_{\rm F}$ for $x \geq 0.79$. This quasi-linear dispersion suggests the existence of the TSS due to the band inversion between the $d_{xz}$ and the $xy^-$ band, which originates from the hybridized $p_{z}$ and $d_{xy}$ orbitals~\cite{kim2024orbital}. The $d_{xz}$ band in FeTe appears as a ``streak" feature at $\overline{\Gamma}$ with no well-defined dispersion (Fig.~\ref{fig2}\textbf{a(VI)}), deviating from the single-particle picture. The orbital-dependent band renormalization near $\overline{\Gamma}$ generally agrees with the measurements on bulk FTS crystals~\cite{liu_experimental_2015, huang_correlation-driven_2022}, and is fundamentally facilitated by the suppression of inter-orbital hopping due to Hund's $J$ coupling~\cite{si2016high}.

% \commentSLY{I will finish the SOC section in the supplementary.}

Near the $\overline{\text{M}}$ point, both a hole pocket with its $k_{\rm F}$ near $\pm 0.1$~\AA$^{-1}$ and a shallow electron pocket with its binding energy $< 20$~meV are observed in 10 UC FeTe$_{0.40}$Se$_{0.60}$ (Fig.~\ref{fig2}\textbf{d(I)}), resembling previous results on multilayer FeSe films~\cite{zhang_distinctive_2016}. For $x \geq 0.60$, only the electron pocket survives, which is assigned to a hybridized $d_{xy/xz}$ band in agreement with measurements on bulk FTS crystals~\cite{yi_observation_2015}. The effective mass of the $d_{xy/xz}$ band rapidly increases when $x$ approaches 1, increasing from $9 \pm 2$~$m_{e}$ at $x = 0.60$ to $39 \pm 9$~$m_{e}$ at $x = 0.93$. Though the $d_{xy/xz}$ band is not strictly parabolic, we use a parabolic function to estimate its effective mass in order to reveal an overall trend of band renormalization. This band renormalization effect can also be seen by plotting the ratio of $k_{\rm F}^2/E_{\rm band}$, where $E_{\rm band}$ represents the electronic bandwidth (Supplementary Fig.~11). For FTS films with a heavy Te content ($x \geq 0.79$), the spectral feature at $E - E_{\rm F} < -0.02$~eV is best characterized as a broad hump in the raw data (Fig.~\ref{fig2}\textbf{c(III)-c(V)}), mimicking the correlation driven hump feature observed in high-temperature superconducting cuprates~\cite{hashimoto_particlehole_2010, chen_incoherent_2019}. Notably, in the limit of $x\sim 1$ the spectral weight of the $d_{xy/xz}$ hybridized band at $\overline{\text{M}}$ is heavily suppressed. Even though the 2D curvature plots at $\overline{\Gamma}$ (Fig.~\ref{fig2}\textbf{b(VI)}) and $\overline{\text{M}}$ (Fig.~\ref{fig2}\textbf{d(VI)}) suggest the existence of $d_{xy}$-derived flat bands, such features can also be due to the large intensity curvature associated with the Fermi edge. The rapid disappearance of the hybridized $d_{xy/xz}$ electron pocket in 10 UC FeTe likely reflects the combined effect of strongly correlated $d_{xy}$ and $d_{xz}$ orbitals near $\overline{\text{M}}$, while the $d_{xy}$ orbital attains the highest renormalization (Fig.~\ref{fig2}\textbf{g}, Supplementary Fig.~12). The effective mass extracted from the innermost hole band near $\overline{\Gamma}$ is heavily influenced by the $d_{xz}$-$xy^{-}$ band inversion, and does not represent the strong correlation effect of the $d_{xz}$ orbital near $\overline{\text{M}}$.

These observations in 10 UC FeTe, together with the systematic evolution of the electronic structures in all FTS films, suggest the occurrence of an orbital selective correlated phase (OSCP) where the $d_{xy}$ band becomes distinctively more correlated than other bands in the limit of $x = 1$~\cite{yi_observation_2015, si2016high, kim2024orbital}. The rapid disappearance of the $d_{xy}$ spectral weight has been attributed to either an orbital-selective Mott phase (OSMP)~\cite{huang_correlation-driven_2022}, or a coherent-to-incoherent transition~\cite{Haule-njp09,Mravlje-prl11,ieki2014evolution}. In the former, the spectral weight of a quasiparticle band is shifted to  Mott-Hubbard bands at higher binding energies defined by the Coulomb $U$~\cite{huang_correlation-driven_2022}. In the latter, the quasiparticle spectral weight still decreases near $E_{\rm F}$, but it can be reformed into a more general incoherent feature~\cite{liu_experimental_2015,damascelli2003angle}. Experimentally, we observe a decrease in the spectral intensity in the Fe $3d$ bands near $E_{\rm F}$ as the Te content increases, and an increase in the spectral intensity in a broad binding energy range of $1-3$~eV (Supplementary Fig.~13). While this may represent evidence for the OSMP, we caution that variation in sample qualities and matrix elements can also contribute to the photoemission intensities. Moreover, finite inter-orbital hopping can lead to an OSMP breakdown at sufficiently low temperatures~\cite{kugler_is_2022}, potentially making our observed strongly correlated regime at finite temperatures a strictly crossover behavior. We thus do not distinguish between these two scenarios, and focus on how OSCP impacts topology and superconductivity.

% \commentSLY{Need to put in the valence band scan in the supplementary in response to Referee 3.}

% Given that the experimental predictions are similar, we will not attempt to distinguish these two scenarios. Here we denote this phenomenon as the OSCP to account for the fact that both effects manifest the strong electronic correlations.

Selected resistance-vs-temperature (R-vs-T) curves for 10 UC FTS thin films with different Te contents are shown in Fig.~\ref{fig2}\textbf{f}. The onset of the superconducting transition is observed for all the samples with Te content ranging from $x = 0.60$ to $x = 1$. The suppression of the superconducting transition due to magnetic fields is confirmed (Supplementary Fig.~15). However, for FeTe the superconducting transition is notably broader and zero resistance is not achieved for temperatures down to $2$~K. Measurements over an extended temperature range show the absence of a clear resistivity drop near 70 K (Supplementary Fig.~16), indicating the suppression of the bicollinear AFM order~\cite{jiang2013distinct}.

\begin{figure}[h]%
\centering
\includegraphics[width=1\textwidth]{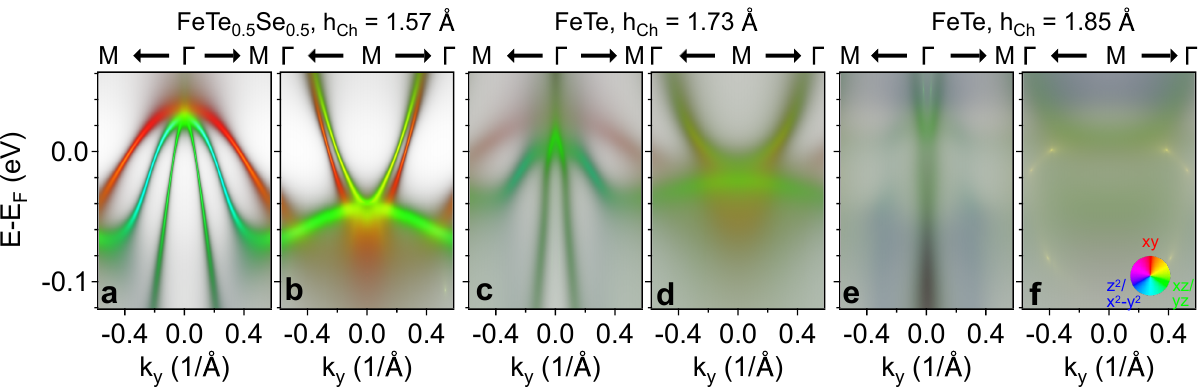}
\caption{\textbf{Calculated spectral functions of FeTe$_{x}$Se$_{1-x}$ using the density functional theory (DFT) in combination with the embedded dynamical mean field theory (eDMFT). a-b,} Spectral functions for bulk FeTe$_{0.5}$Se$_{0.5}$ near $\Gamma$ (\textbf{a}) and near $M$ (\textbf{b}), with the in-plane lattice constant fixed to $3.88$~\AA~and the chalcogen height determined by DFT+eDMFT. \textbf{c-d,} Same as panels \textbf{a} and \textbf{b} for FeTe. \textbf{e-f} Spectral functions of FeTe, but with the Te height manually fixed at $1.85$~\AA.}\label{fig3}
\end{figure}

We perform first-principles calculations using density functional theory combined with the embedded dynamical mean field theory (DFT + eDMFT) to reveal the systematic evolution of electronic correlations with an increasing Te content. %To mimic the substrate-induced straining effect as confirmed by scanning transmission electron microscopy (STEM) (Supplementary Fig. 6), we first calculate the spectral functions for bulk FeTe$_{0.5}$Se$_{0.5}$ (Fig.~\ref{fig3}\textbf{a}, \ref{fig3}\textbf{b}) and FeTe (Fig.~\ref{fig3}\textbf{c}, \ref{fig3}\textbf{d}) with their in-plane lattice constants fixed to $3.9$~\AA. 
We first calculate the spectral functions for bulk FeTe$_{0.5}$Se$_{0.5}$ (Fig.~\ref{fig3}\textbf{a}, \ref{fig3}\textbf{b}) and FeTe (Fig.~\ref{fig3}\textbf{c}, \ref{fig3}\textbf{d}) using an in-plane lattice constant of 3.88~\AA, as determined by scanning transmission electron microscopy (STEM) (Supplementary Fig.~18). Importantly, it is well-known that the experimental $k_{z}$ dependence of the FTS electronic structure is much weaker than any first-principles predictions~\cite{li2024orbital, lohani2020band}. Moreover, FTS compounds with different Te:Se ratios may not have the same inner potential for $k_{z}$ determination~\cite{liu_experimental_2015}. These complications make it challenging to determine the $k_{z}$ for theory-experiment comparison (Supplementary Note 12). Here we present theoretical calculations for $k_{z} = 0$, considering the successful reproduction of the spectral function in the strongly correlated limit. The eDMFT-optimized average chalcogen heights are $1.57$~\AA ~for FeTe$_{0.5}$Se$_{0.5}$ and $1.73$~\AA ~for FeTe. For the $d_{xy}$ band, the calculated effective mass increases from 15.56 $m_e$ to 35.20 $m_e$ going from FeTe$_{0.5}$Se$_{0.5}$ to FeTe. Even with the increased electronic correlation for the $d_{xy}$ orbital, the calculated spectral function for strained FeTe (Fig.~\ref{fig3}\textbf{c}, \ref{fig3}\textbf{d}) cannot completely capture the corresponding ARPES spectra for $x = 1$ (Fig.~\ref{fig2}\textbf{a(VI)}, \ref{fig2}\textbf{c(VI)}). While chalcogen height directly influences electronic correlations in FTS, it is not the only factor; effects such as polaronic coupling~\cite{liu2013measurement} or substrate-induced electric field are not accounted for in the DMFT calculations. To further explore the potential impact of stronger electronic correlations, we perform DFT+eDMFT calculations on FeTe with a manually increased chalcogen height of $1.85$~\AA. As shown in Fig.~\ref{fig3}\textbf{e} and \ref{fig3}\textbf{f}, the resulting spectral function of FeTe reproduces well the experimental observations: the $d_{xy}$ band at $\Gamma$ and the $d_{xy/xz}$ band at $\text{M}$ become incoherent upon entering the OSCP; the $d_{xz}$ band at $\Gamma$ is characterized by an incoherent ``streak'' feature. %For ultrathin films, $k_{z}$ in ARPES may not be easily determined due to the 3D-2D crossover~\cite{szalowski2006critical}. We note that the strong correlation effect is insensitive to the exact value of $k_{z}$~\cite{liu2015experimental}. 
%For simplicity we did not incorporate the SOC for the calculations presented in Fig.~\ref{fig3}. 
For simplicity we do not incorporate the SOC for the calculations presented in Fig.~\ref{fig3}. Additional calculations incorporating the SOC demonstrate that the band dispersions in the FeTe limit are minimally affected (Supplementary Fig.~9). Our DFT+eDMFT calculations reveal that the general evolution of the spectral function when tuning the Te:Se ratio is consistent with the experimental observation, in that the $d_{xy}$ orbital becomes selectively more correlated.

% \commentSLY{I need to add the supplementary note on the kz discussion. I need to add the supplementary note on SOC.}

\begin{figure}[h]%
\centering
\includegraphics[width=1\textwidth]{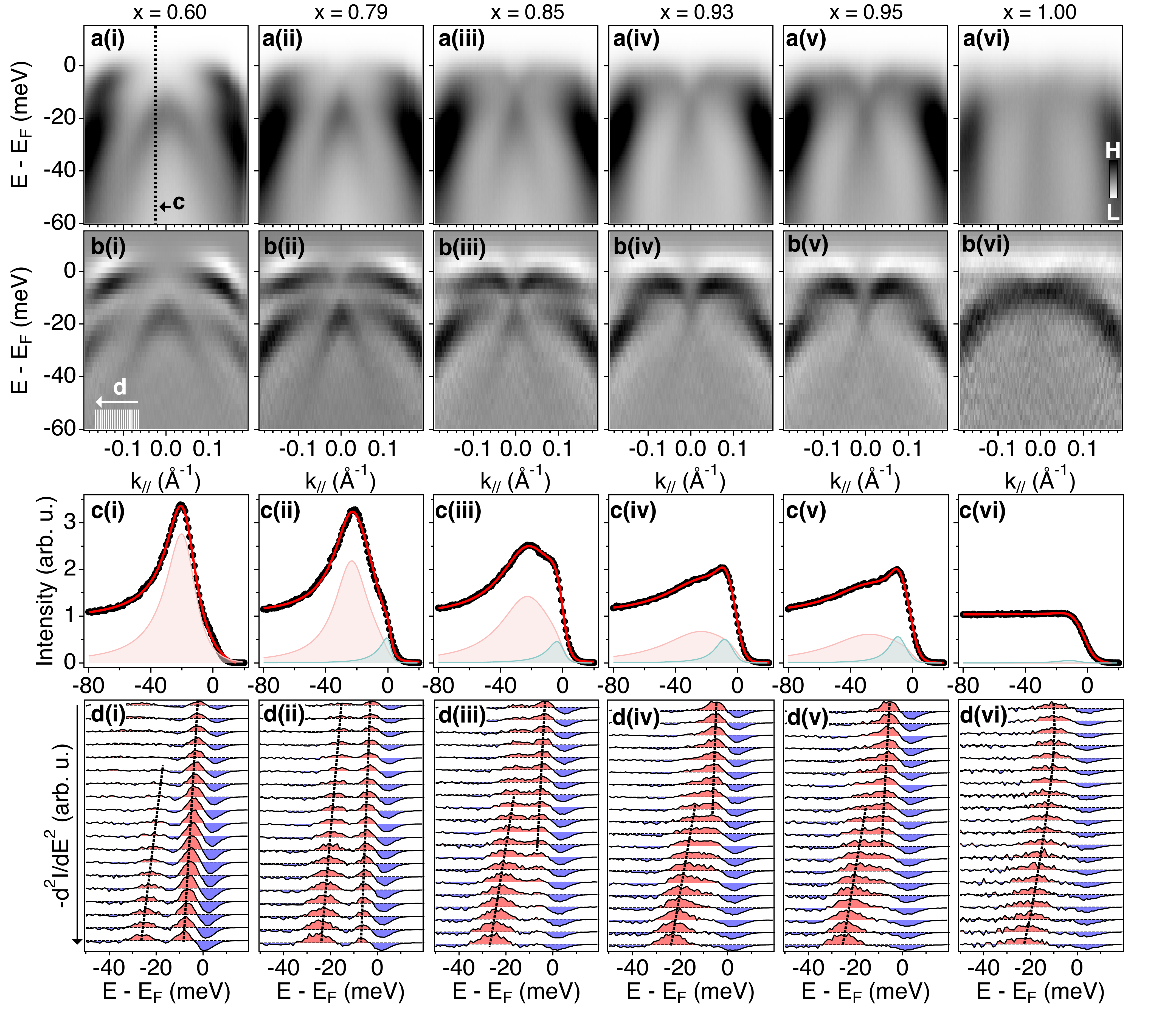}
\caption{\textbf{Loss of quasiparticle coherence for FeTe$_x$Se$_{1-x}$ toward the $x = 1$ limit. a(i)-a(vi),} Evolution of the ARPES spectra as a function of the Te content ($x$) near the $\overline{\Gamma}$ point using $p$-polarized 6 eV laser-based ARPES. Measurement temperature was at $20$~K. \textbf{b(i)-b(vi),} Corresponding second derivative plots of the raw data in \textbf{a}. \textbf{c(i)-c(vi),} Black circles are EDCs taken at $-0.025$~\AA$^{-1}$. Red curves are fits to the EDCs. Red and green shaded areas represent the components corresponding to the lower and upper Dirac cones, respectively. \textbf{d(i)-d(vi),} Inverted second derivative curves corresponding to momentum-dependent EDCs in the range between $-0.065$ and $-0.165$~\AA$^{-1}$.}\label{fig4}
\end{figure}

%%%%%=============================================================================%%%%
%%%%  Caption for original Fig3.
%%%%%=============================================================================%%%%
%\caption{\textbf{Calculated single-particle spectral functions of FeTe$_{x}$Se$_{1-x}$ using the density functional theory (DFT) in combination with the embedded dynamical mean field theory (eDMFT). a,} Calculated spectral function for bulk FeTe$_{0.5}$Se$_{0.5}$ near the $\Gamma$ point, with its in-plane lattice constants set to the same value as those for the SrTiO$_{3}$(100) substrate. \textbf{b,} Same as \textbf{a} but for bulk FeTe, with the in-plane lattice constants also strained by the SrTiO$_{3}$ substrate. \textbf{c,} Same as \textbf{b} but for unstrained bulk FeTe$_{0.9375}$Se$_{0.0625}$. \textbf{d,} Same as \textbf{c} but for unstrained bulk FeTe. [\textcolor{red}{\textbf{Place Holder}: to be replaced by strained $\text{FeTe}$ but with a manually increased $\text{h}_{\text{ch}}$}] \textbf{e-h}, Calculated spectral functions near the $M$ point for (\textbf{e}) substrate-strained FeTe$_{0.5}$Se$_{0.5}$, (\textbf{f}) substrate-strained FeTe, (\textbf{g}) unstrained FeTe$_{0.9375}$Se$_{0.0625}$, and (\textbf{h}) unstrained FeTe, respectively.}\label{fig3}

We employ $6$~eV laser-based ARPES to directly visualize the TSS in FTS thin films. The $p$-polarized laser allows a nonzero sensitivity to all the $t_{2g}$ orbitals and the chalcogen $p_{z}$ orbital (Supplementary Note~2). For x = 0.60, no TSS is observed and the FTS film is topologically trivial. As the Te content increases to 0.79, a crossing feature near $-10$~meV emerges. This crossing feature is not fully evident for $x=0.79$ using $p$-polarized light, but is clearly resolved using $s$-polarized light (Supplementary Fig.~4h). The critical doping for the topological phase transition is thus between $0.60$ and $0.79$, which is different from $x\approx 0.5$ in FTS bulk crystals~\cite{wang_topological_2015, wu_topological_2016, xu_topological_2016, zhang_observation_2018, wang_evidence_2018}. This modified critical doping is consistent with the observation on monolayer FTS films~\cite{shi_fete1se_2017,peng2019observation}. 

% Importantly, our 10UC FTS films are epitaxially strained to the STO substrates as determined by STEM (Supplementary Fig. 6). The in-plane tensile strain generally leads to a lower average chalcogen height. As a result, a higher Te content is required to reach the critical chalcogen height needed to drive the topological band inversion. This is consistent with our observation that the average Te height ($1.74$~\AA) in 10 UC FeTe (Supplementary Fig. 7) is smaller than that for bulk FeTe ($1.765$~\AA)~\cite{viennois_effect_2010}.

In the heavy Te regime ($x > 0.93$), the correlation effect starts to influence the topological electronic structure. The lower Dirac cones are increasingly smeared out, with its spectral weight systematically suppressed as a function of $x$ (Fig.~\ref{fig4}\textbf{c}, Supplementary Fig.~21). At $x = 1$, the TSS is unidentifiable, leaving a ``vertical streak" feature consistent with the 21.2~eV ARPES data (Fig.~\ref{fig2}\textbf{a(VI)}). Instead, a small, faint electron pocket becomes visible under $s$-polarized photons (Supplementary Fig.~4\textbf{f}). This electron pocket was predicted by a tight-binding model under the condition of complete $d_{xz}$-$xy^{-}$ band inversion throughout the $\Gamma$-Z trajectory (Supplementary Note~14). The nonexistence of a sharp TSS and the existence of the ``vertical streak" in the FeTe limit cannot be reconciled with the single-particle picture~\cite{wang_topological_2015}, where the presence of the TSS is only determined by the $p_z$ - $d_{xz}$ band inversion and should not exhibit such a strong doping dependence. Moreover, the hybridization gap between the $d_{xy}$ and $d_{yz}$ bands, as resolved by second-derivative plots of the EDCs, is clearly visible for lower $x$ values (Fig.~\ref{fig4}\textbf{d(i)-d(v)}), but rapidly diminishes as $x$ approaches 1 (Fig.~\ref{fig4}\textbf{d(vi)}, Supplementary Fig.~22). For 10 UC FeTe the $d_{yz}$ band does not exhibit any hybridization gap. The lack of $d_{xy}$-$d_{yz}$ hybridization in FeTe is also reproduced by our DMFT calculations incorporating the SOC (Supplementary Note~5).

% \commentSLY{I will add a supplementary note on the electron pocket in FeTe.}

\begin{figure}[h]%
\centering
\includegraphics[width=1\textwidth]{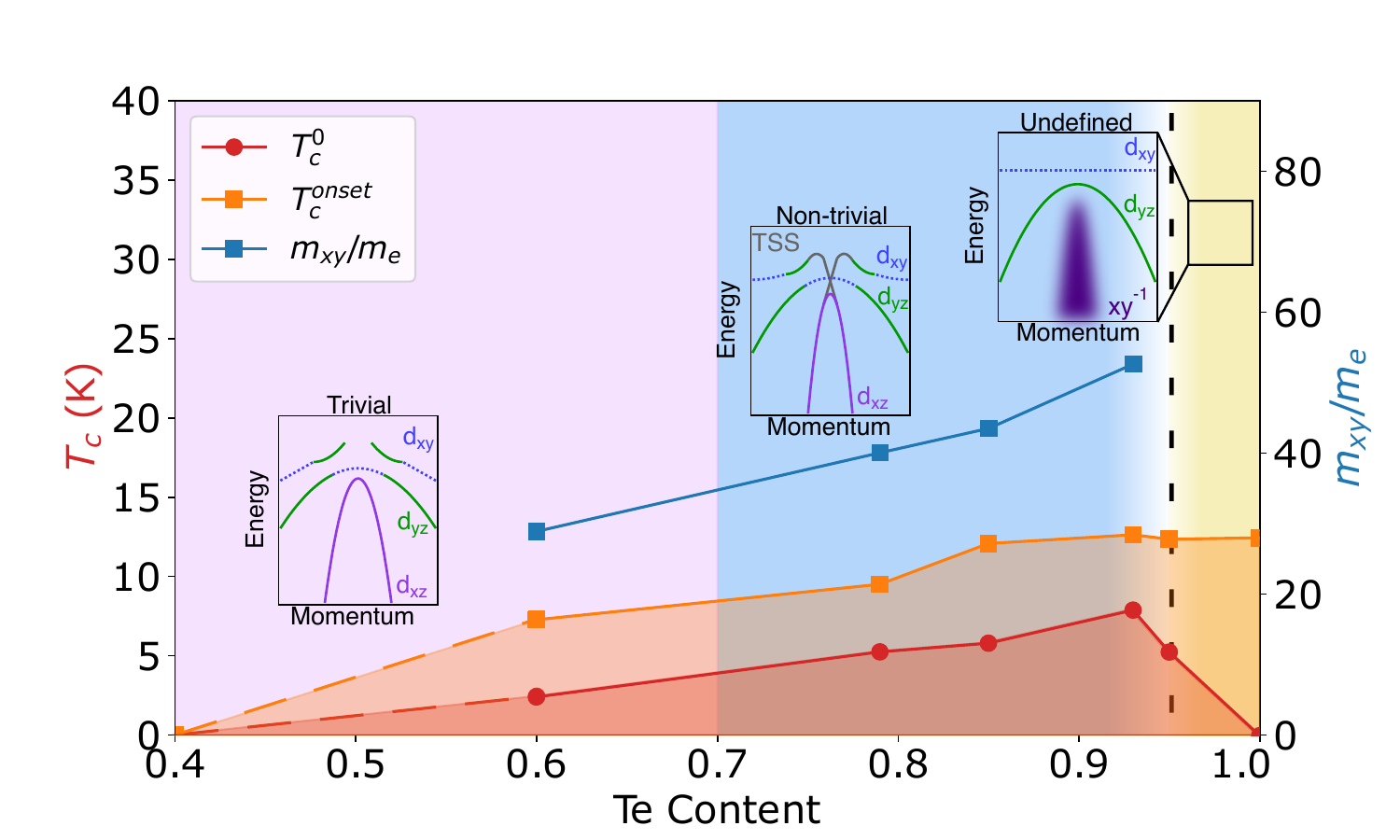}
\caption{\textbf{Topological phase diagram of 10 UC FeTe$_{x}$Se$_{1-x}$ thin films.} The onset transition temperature ($T_c^{\text{onset}}$) and the temperature ($T_c^0$) at which the resistance reaches 1\% of the normal state resistance at 20 K are plotted against Te content, $x$. The blue curve shows the effective mass of the $d_{xy}$ band as a function of $x$. Near the FeTe limit, an undefined phase emerges, characterized by smeared topological surface states originating from localized electrons within the OSCP. Concurrently, this region also exhibits a suppression of superconductivity.}\label{fig5}
\end{figure}

\section{Discussion}\label{sec12}

%The ``undefined" regime reflects orbital-selective correlation induced localization at finite temperatures that obscures the TSS. Due to measurement limitations we cannot determine whether it represents a quantum phase in the $T = 0$ limit.

%We note that the ``undefined" regime shown in Fig.~\ref{fig5} does not represent a quantum phase in the $T = 0$ limit; rather,

Our results illustrate the profound impact of strong correlations on the band topology in FTS (Fig.~\ref{fig5}). The ``undefined" regime reflects orbital-selective correlation-induced localization at finite temperatures that obscures the TSS. Due to measurement limitations we cannot determine whether it represents a quantum phase in the $T = 0$ limit. 
%Similar to the case of a Mott insulating phase, one may identify the inverse quasiparticle weight ($Z^{-1}$) as an order parameter for the OSCP (Supplementary Fig.~12). %\textcolor{red}{[Haoran: Not anymore:. We note that even though the measurement of an expected $1$-$2$~meV superconducting gap on the TSS is beyond our current instrument resolution (Methods), we infer from the existing literature on bulk FTS~\cite{zhang_observation_2018} that the TSS is proximity-coupled to the bulk superconductivity, thus leading to a topological superconducting state.]} 
By systematically varying the Te content $x$, we reveal the rapid shift from the itinerant regime for $x < 0.95$ to the near-localized regime for $d_{xy}$ electrons when $x$ approaches 1. The reduction of the $d_{xy}$-$d_{yz}$ hybridization can be reconciled with the diminishing spectral weight of the $d_{xy}$ band. In the OSCP picture, the $d_{xy}$ spectral weight is shifted to higher binding energies and hence does not hybridize with the $d_{yz}$ band (Supplementary Fig.~13). In the meantime, the $d_{xz}$ band attains the $xy^-$ character spanning an extended energy range $> 50 $~meV due to the band inversion (Supplementary Fig.~25)~\cite{kim2024orbital,peng2019observation}. Since the $xy^-$ band has contributions from the $d_{xy}$ orbital, the localization of the $d_{xy}$ orbital leads to substantial energy broadening, thus smearing out both the original $d_{xz}$ band and its derived TSS. Considering the strong renormalization of the $xy^-$ band, our interpretation also aligns with the negligible $k_{z}$ dependence of the $d_{xz}$ band top~\cite{lohani2020band, li2024orbital} (Supplementary Fig.~26).

% Additional data taken by systematically tuning the thickness of FeTe$_{0.85}$Se$_{0.15}$ shows that the OSCP is also achieved in the monolayer limit, due to the substrate-enhanced electronic correlations~\cite{Mandal:2017} (Supplementary Fig. 12). Scanning tunneling microscope (STM) confirmed the sample quality down to the monolayer limit (Supplementary Fig. 13). Remarkably, the disappearance of the $d_{xy}$ spectral weight and the energy smearing of the $d_{xz}$ band are always correlated when tuning either the Te:Se ratio or the film thickness. This corroborates the picture of $d_{xy}$-driven OSCP in understanding the energy smearing in the $d_{xz}$ band. Interestingly, it also rules out the $k_{z}$ integration as a possible scenario to explain the energy smearing - the electronic structure of 1UC FeTe$_{0.85}$Se$_{0.15}$ does not disperse along $k_{z}$.

Notably, a previous DMFT study predicted that the FTS system becomes topologically trivial upon entering the OSCP regime~\cite{kim2024orbital}. In our experiment, we find it challenging to distinguish the trivial phase from an undefined phase where the $xy^-$-$d_{xz}$ band inversion still occurs, but the inverted $xy^-$ band has a substantially enhanced self energy, resulting in a smeared out TSS. A heavily renormalized electronic state can be represented as $\Psi(r, t)$ = $\Psi(r){\rm e}^{(-i E_{B} t / \hbar)}{\rm e}^{(-\text{Im}(\Sigma) t / \hbar)}$. When the imaginary part of the self energy $\text{Im}(\Sigma)$ exceeds the electronic binding energy $E_{B}$, the quasiparticle is not well-defined~\cite{Grimvall1981TheEI}. As topology in FTS is defined by the symmetries of quasiparticle bands~\cite{xu_topological_2016}, the system is effectively topologically trivial due to the poorly defined quasiparticles. This correlation-driven change of the band topology goes much deeper than simply modifying the shape of topological bands. It is beyond traditional DFT calculations (Supplementary Fig.~27), and highlights the critical need to incorporate electronic correlations by performing DFT+eDMFT calculations (Supplementary Figs.~30-34). Near $x = 1$, the few-eV scale correlation physics dominates the few-meV scale topological physics, resulting in fundamental changes of the topological properties. %\commentSLY{I added something to address Referee 2's general concerns: if simply modifying the shape of bands, then it's not interesting.}

Our results highlight that FTS thin films are characterized by a different phase diagram as compared to their bulk counterpart. In bulk FTS, superconductivity is fully suppressed for $0.92 < x < 1$ and a bi-collinear AFM phase emerges~\cite{hu_coupling_2013, huang_correlation-driven_2022}; %However, previous transport studies on FTS thin films with thicknesses $< 100$~nm~\cite{han2010superconductivity, sato2024molecular} reported that the AFM phase in this doping range is hampered and the superconducting dome extends to $x \approx 1$, in agreement with our results (Supplementary Fig. 5). The unique superconducting dome for FTS thin films can be understood by a two-fold mechanism. First, %the structural modification in FTS thin films, as compared to the bulk counterparts, leads to electronic changes. 
the topological transition occurs near $x\sim 0.5$~\cite{zhang_observation_2018,zhang_multiple_2019}. In our FTS films, the superconducting dome extends to $x\sim 1$; the AFM phase is suppressed; the topological transition is shifted to $x\sim 0.7$. The first two observations are consistent with previous transport studies on FTS thin films~\cite{han2010superconductivity, sato2024molecular}, which were attributed to the suppression of the magnetic and lattice symmetry breaking due to substrate straining. The superconducting dome is also consistent with our superconducting gap measurements at $x = 0.60$ and $x = 1.00$ (Supplementary Note~16). The shift of the topological transition and that of the superconducting dome may be generally understood by considering the reduction of the average chalcogen height due to the in-plane tensile strain (Supplementary Fig.~18). For instance, the average Te height of 10 UC FeTe is $1.74$~\AA~(Supplementary Fig.~19), which is smaller than $1.765$~\AA~in bulk FeTe~\cite{viennois_effect_2010}. On the other hand, STEM measurements yield asymmetric chalcogen heights above and below the Fe sheet (Supplementary Fig.~19), breaking the glide-mirror symmetry. %\commentSLY{The discussion on the lack of band splitting due to glide-mirror symmetry breaking is off the main topic. I moved it to Supplementary Note~4.}%Such a structure breaks the glide-mirror symmetry~\cite{zhang2016superconducting} and should result in a splitting of the electron pocket near the $\overline{M}$ point. However, previous studies on monolayer FeSe~\cite{zhang2016superconducting}, which also breaks the glide-mirror symmetry, suggested that such a splitting can be negligible ($<$ 5 meV) near the electron pocket bottom. 
Yet, even when this symmetry is broken, the $\mathbb{Z}_2$ topological invariant remains unchanged (Supplementary Fig.~28). The superconducting phase dome of our 10 UC FTS films is also notably different from that of monolayer films~\cite{li_interface-enhanced_2015, liu_spatial_2023}, which likely arises from the monolayer films' unique electronic interactions present at the film-substrate interface.

On the other hand, the structural modification is not the only factor dictating the phase diagram for ultrathin FTS films. The structural similarity between FeTe$_{0.93}$Se$_{0.07}$ and FeTe (Supplementary Fig.~19) contrasts their substantial differences in the electronic spectra and superconducting transitions (Fig.~\ref{fig2} and~\ref{fig4}). Here the rapid chemical doping dependence is analogous to that of cuprates~\cite{Shen2004}: a few percent of Se doping introduces mobile $d_{xy}$ carriers, which subsequently collapses the OSCP. %Second, the change in the $d_{xy}$-derived electronic bands both at $\overline{\Gamma}$ and $\overline{M}$ directly influences superconductivity. 
The doping-dependent variation of the superconducting transition represents the profound impact of strong correlations on superconductivity. Consider a generic electron-boson coupling strength $\lambda \sim N(E_{\rm F}) D^{2}$, where $N(E_{\rm F})$ is the electronic density of states at $E_{\rm F}$ and $D$ is a generalized coupling potential. When $x$ increases from $0.60$ to $\sim 0.93$, the $d_{xy}$-derived bands are flattened both at $\overline{\Gamma}$ and $\overline{\text{M}}$, contributing a rapidly increasing $N(E_{\rm F})$. Moreover, $D$ for the case of electron-phonon coupling can also be enhanced by electronic correlations~\cite{mandal2014strong,gerber2017femtosecond}. These factors lead to the enhancement of superconductivity. When $x$ increases from $0.93$ to $1$, the spectral weight of $d_{xy}$-derived bands quickly diminishes due to OSCP, leading to an effectively reduced $N(E_{\rm F})$. This likely gives rise to the suppression of a coherent superconducting state.

Our work demonstrates that strong correlations, superconductivity and nontrivial topology in epitaxially grown FTS thin films are no longer separate entities. In fact, it is feasible to tune the topological superconducting phase by modulating the electron-electron interactions. This can be achieved by engineering the epitaxial strain or by changing the dielectric environment. For instance, creating an STO/FTS/STO sandwich structure~\cite{coh2016proposal} can fully contain the Coulomb interactions within a high-dielectric-constant medium, leading to strongly modified electron-electron and electron-phonon interactions and potentially high-temperature topological superconductivity.

\backmatter
\pagebreak

\section{Methods}
\subsection{Sample growth}

For the ARPES measurements, 0.05\% wt Nb-doped SrTiO$_3$(100) substrates from SHINKOSHA CO., LTD were used. They were cut into $10$ mm $\times$ $2.5$ mm rectangles. After ultrasonic cleaning in Acetone and Isopropyl Alcohol, the substrates were annealed in ultra-high vacuum at $\sim$1000 $^\circ$C for about 30 minutes. The FeTe$_{x}$Se$_{1-x}$ thin films were synthesized at a substrate temperature of 270 $^\circ$C by co-evaporating Fe, Te, and Se in an MBE chamber. For all the FeTe$_{x}$Se$_{1-x}$ thin films, the temperature of Fe and Te were kept at 1180 $^\circ$C and 265 $^\circ$C, respectively. Te content was controlled by varying the Se temperature from 110 $^\circ$C to 135 $^\circ$C. All films were post-annealed at the growth temperature for 1 hour before ARPES measurement. The growth rate and sample quality were calibrated by scanning tunneling microscopy (STM) measurements (Supplementary Fig.~37). \\
For the transport measurements, insulating SrTiO$_3$(100) substrates from SHINKOSHA CO., LTD were used. They were cut into $10$ mm $\times$ $5$ mm rectangles. After ultrasonic cleaning in Acetone and Isopropyl Alcohol, the substrates were etched in buffered oxide etch (BOE) for 1 minute and were then annealed at 1000~\textcelsius~with an O$_{2}$ flow of 0.5 L/min for 3 hours. 

\subsection{ARPES measurements}

ARPES measurements were typically carried out at 20~K on the multi-resolution photoemission spectroscopy platform established at the University of Chicago~\cite{yan_integrated_2021}. Helium-lamp-based ARPES measurements were performed using 21.2 eV He-I$\alpha$ light. Laser-based ARPES measurements were performed with a spatial resolution of 10 $\times$ \SI{15}{\micro\meter}$^2$ and a typical energy resolution of $\sim$4 meV using 206-nm pulses with a repetition rate of 80 MHz. In the superconducting gap measurements, the energy resolution was further optimized to $\sim 2.4$~meV and the sample temperature was varied between $8$ and $25$~K.

\subsection{STEM measurements}

FeTe$_{x}$Se$_{1-x}$ samples with Te capping were prepared for cross-sectional STEM imaging using standard focused ion beam (FIB) lift-out procedures in a Thermo Fisher Scientific Helios 600i DualBeam FIB-SEM. A cryo-can was used during the thinning process to reduce redeposition. These FIB cross-sectional samples were stored in an inert atmosphere glovebox between sample preparation and STEM imaging. FeTe$_{x}$Se$_{1-x}$ cross sectional samples were imaged in a Thermo Fisher Scientific Themis Z aberration-corrected STEM operated at 300 kV with a convergence angle of 25.2 mrad. Scale for each atomic resolution image was calibrated with the bulk STO spacing of the sample substrate.

\subsection{STM measurements}

Samples were transferred from the University of Chicago to the Center for Nanoscale Materials utilizing a homebuilt vacuum suitcase. The vacuum suitcase was compatible with the Omicron VT scanning tunneling microscope (STM), which allowed for characterization of samples that remained in ultrahigh vacuum conditions after initial growth. Imaging was done with a standard electrochemically etched W tip. A topographic image of 0.8 UC FTS is illustrated in Supplementary Fig.~37. Sample bias of +2.0 V and a tunneling current setpoint of 100 pA were used during the scanning.

\subsection{DFT+eDMFT calculations}
We used a fully self-consistent DFT+eDMFT implementation~\cite{eDMFT2010,eDMFT2018}, where the charge density, impurity level, chemical potential, self-energy, and the lattice and impurity Green's Function were computed self-consistently. The DFT+eDMFT functional had a form of the exact Klein functional~\cite{Klein,Kotliar-rmp06}, with the approximation of the correlation self-energy to be truncated to the local part of each correlated atom in the unit cell~\cite{Kotliar-rmp06}. In addition, the less correlated chalcogen atoms and interstitial charge were treated on the DFT level. The double-counting between DFT and DMFT was subtracted exactly~\cite{ExactDC:2015}.
The resulting non-perturbative self-energy, which the Klein functional requires, was in practice calculated by solving the quantum impurity problem in the presence of a self-consistent electronic bath (mean-field environment). The imaginary axis self-energy was calculated using the continuous-time quantum Monte Carlo method from local properties of the Fe ion and was added to the DFT Kohn-Sham Hamiltonian~\cite{Kotliar-rmp06}. To represent the lattice problem, we used the WIEN2k package,~\cite{wien2k} which uses the full potential augmented plane wave method. To obtain the local Green's function, required by the impurity solver, we used projection to the very localized orbitals contained within the muffin-tin spheres of correlated atoms~\cite{eDMFT2010}. This projection/embedding was done on correlated orbitals within $\pm$ 10 eV of the Fermi-energy, which can capture spectral weight of all electrons in the solid.  A Monkhorst-Pack k-point mesh of 15$\times$15$\times$10 and the Local Density Approximation (LDA) exchange-correlation was employed at the DFT level, while at the DFT+eDMFT level we used a total of 5 million Monte Carlo steps, a Coulomb's interaction (U) of 5.0 eV, a Hund's coupling (J$_{H}$) of 0.8 eV, and a temperature of $116$~K (100 $\beta$) to model the effects of alloying tellurium and selenium on the FTS system. The values of $U$ and $J$ were computed using the constrained-DMFT method and had been used previously and agreed with experiments~\cite{Yin-NatMat,Mandal:2017}. 
To represent 10 UC films, we modeled the FeTe$_{0.5}$Se$_{0.5}$ and FeTe systems in their bulk phases. FeTe$_{0.5}$Se$_{0.5}$ was modeled by having separate sheets of selenium and tellurium. The in-plane lattice parameters of FeTe$_{0.5}$Se$_{0.5}$ and FeTe were set to the experimental value of 3.88~\AA. For FeTe$_{0.5}$Se$_{0.5}$, we optimized the Se/Te positions using eDMFT, which incorporated the effects of the electron's entropy~\cite{Forces,FreeE:2015}. For FeTe, we performed calculations using both the eDMFT-optimized Te height of 1.73~\AA, and a manually tuned height of 1.85~\AA~to investigate the strong correlation effect. We used the maximum entropy method to analytically continue the self-energy from the imaginary to the real axis. We then computed the spectral functions [A(k,$\omega$)] shown in Fig. 3a-3f and Supplementary Fig.~30-34 using color to resolve the orbital characters of the bands near $E_{\rm F}$. Calculations incorporating the SOC are shown in Supplementary Fig.~8 and~9.

%To explore the effects on the electron correlations due to straining and alloying we calculate all five Fe-3$d$ orbitals' spectral weights Z (inverse of mass enhancement $m^{*} \backslash m_{\rm DFT}$) using $Z=(1-\frac{d\Sigma(i\omega))}{d\omega} |_{\omega \rightarrow 0})^{-1}$.

%Here we consider the in-plane lattice parameter of STO, which is 3.905\AA \ and optimize the chalcogen positions. We use the maximum entropy method to analytically continue the self-energy from the imaginary to the real axis.

\subsection{Tight-binding calculations}
We adopted the eight-band effective model to simulate the band structure of multi-layer FTS thin films in the topologically nontrivial regime~\cite{zhang_multiple_2019}. The adopted onsite energy of the $d_{xy}$ orbital was modified to $m_{0}^{3}=0.001$~eV and the out-of-plane lattice constant was $6.27$~\AA.

\subsection{Transport measurements}

10 UC FTS thin films were grown using the recipe described in the Sample growth section. A capping layer was deposited by keeping the Te flux for $\sim$5 minutes after film growth. Electrical contacts for transport measurements were established using small indium dots, which were manually applied to the samples. The average distance between adjacent dots was $\sim 0.5$~mm. Electrical transport measurements were conducted using a Physical Property Measurement System (PPMS). An excitation current of approximately \SI{1}{\micro\ampere} was applied for the resistivity measurements.

\subsection{Inductively Coupled Plasma Mass Spectrometry (ICP-MS)}

For acid digestion, each sample was treated with \SI{750}{\micro\liter} of hydrochloric acid (HCl, $\sim 38$~wt\%) and \SI{250}{\micro\liter} of nitric acid (HNO$_{3}$, $\sim 70$~wt\%), then left for at least three days to ensure complete dissolution. The resulting supernatant was then diluted with a $3$\% nitric acid solution for subsequent ICP-MS analysis. All measurements were performed using either the Thermo iCAP Q ICP-MS or Thermo iCAP RQ ICP-MS. The calibration curves demonstrated excellent linearity, with coefficients of determination (R$^{2}$) of at least 0.9999 for all elements of interest, ensuring high analytical accuracy and precision.

\bmhead{Acknowledgments}

We thank Zhi-Xun Shen and Rafael Fernandes for helpful discussions. MBE and ARPES measurements were supported by NSF via Grant No. DMR-2145373. Transport measurements were done at facilities supported by NSF via Grant No. DMR-2011854. Fabrication of electrical contacts for transport measurements was supported by NSF via Grant CMMI-2240489. S.M. and C.L.J. acknowledge the support from the Air Force Office of Scientific Research by the Department of Defense under Award No. FA9550-23-1-0498 of the DEPSCoR program. S.M. and C.L.J.  benefited from the Frontera supercomputer at the Texas Advanced Computing Center (TACC)
at The University of Texas at Austin, which is supported by National Science Foundation Grant No. OAC-1818253. S.M. also acknowledges the support from NSF OAC-2311558. STEM measurements were supported by the Air Force Office of Scientific Research under award number FA9550-20-1-0302. STEM measurements were carried out in part in the Materials Research Laboratory Central Facilities at the University of Illinois at Urbana-Champaign. ICP-MS measurements were supported by the U.S. DOE Basic Energy Sciences under Grant No. DE-SC0023317. X. W. acknowledges support from the National Key R\&D Program of China (Grant No. 2023YFA1407300) and the National Natural Science Foundation of China (Grant No. 12447103). STM measurements performed at the Center for Nanoscale Materials, a U.S. Department of Energy Office of Science User Facility, was supported by the U.S. DOE, Office of Basic Energy Sciences, under Contract No. DE-AC02-06CH11357. 

%\textcolor{red}{Need funding acknowledgements from Pinshane and Nathan.}

\bmhead{Data availability}
All data that support the plots within this paper and other findings of
this study are available from the corresponding author. Source data are provided with this manuscript.

\bmhead{Code availability}
The code that supports the findings of this study is available from the corresponding author.

\bmhead{Competing interests}
The authors declare no competing interests.

\bmhead{Author contributions}
H.L., C.Y., and S.Y. conceived and designed the experiment. H.L., C.Y., Q.G., G.B., and K.D.N. grew the thin films and performed the ARPES experiment. H.L., P.S., and Y.B. performed the electrical transport measurements. G.M.N. and P.Y.H. performed the STEM measurement. N.P.G., C.Y. and H.L. performed the STM measurement. C.J. and S.M. performed the DFT+eDMFT calculation. X.W. and C.-X.L. performed the tight-binding calculation. G.Y., S.C, and C.L. performed the ICP-MS measurements. H.L. and S.Y. analyzed and interpreted the experimental data. All authors participated in discussions and in writing of the manuscript.

\end{document}